\documentclass[11pt]{article}
\usepackage{graphicx}
\usepackage{subfloat}
\DeclareGraphicsExtensions{.pdf}
\usepackage{amsmath}
\usepackage{amsfonts}   
\usepackage{amssymb}

\begin{document}
\date{}
\title{{\bf{\Large Critical phenomena in Born-Infeld AdS black holes}}}
\author{
{\bf {{\normalsize Rabin Banerjee}}$
$\thanks{e-mail: rabin@bose.res.in}}
,$~${\bf {{\normalsize Dibakar Roychowdhury}}$
$\thanks{e-mail: dibakar@bose.res.in}}\\
 {\normalsize S.~N.~Bose National Centre for Basic Sciences,}
\\{\normalsize JD Block, Sector III, Salt Lake, Kolkata-700098, India}
\\[0.3cm]}



\maketitle


\begin{abstract}
We investigate the thermodynamics of critical phenomena in Born-Infeld AdS black holes using a canonical ensemble. The critical behavior has been studied near the critical point which is characterized by a discontinuity in the heat capacity at constant charge. We explicitly calculate the critical exponents of the relevant thermodynamic quantities. These exponents satisfy all the thermodynamic scaling laws. We also check the Generalized Homogeneous Function (GHF) hypothesis (or the scaling hypothesis) which is shown to be compatible with the thermodynamic scaling laws. Finally we check the validity of the scaling laws of second kind which include the critical exponents associated with the spatial correlation. In the appropriate limit our results also provide the corresponding expressions for the Reissner Nordstrom AdS case.    
\end{abstract}

\section{Introduction}
Thermodynamics of black holes has been a fascinating topic of research since the discovery of a remarkable mathematical analogy between the laws of thermodynamics and the laws of black hole mechanics derived from general relativity \cite{bch}. Following the formal replacement $ E\rightarrow M $, $ T\rightarrow c\kappa $ and $ S\rightarrow \frac{A}{8 \pi c} $ in the laws of thermodynamics, the corresponding laws governing the black hole mechanics emerge in a natural way \cite{bk1}, where $ A $ is the area of the event horizon, $ \kappa $ is the surface gravity and $ c $ is a constant \footnote{For an alternative development based on exact differentials, see \cite{rb}.}. Once black holes are identified as thermodynamic objects, it is natural to study various thermodynamic properties associated with them. An important issue in this context is the phase transition phenomena in black holes which was first studied in \cite{dav1}.

In ordinary thermodynamics, phase transition phenomena plays an important role to explore thermodynamic properties of various systems. The main objective in the theory of phase transition is to study the behavior of a given system in the neighborhood of the critical point, which is characterized by a discontinuity in some thermodynamic variable. In usual thermodynamics various physical quantities pertaining to a system suffer from a singularity near the critical point. The effect of diverging correlation length near the critical point manifests as divergences of these thermodynamic quantities. This is known as \textit{static critical phenomena}. It is customary in usual (equilibrium) thermodynamics to express all these singularities by a set of static critical exponents \cite{stanley2}, which determine the qualitative nature of the critical behavior of a given system near the critical point. The effect of diverging correlation length in case of non equilibrium thermodynamics results in a critical slowing down near the critical point which is parametrized by the \textit{dynamic critical exponents}, and these two sets of exponents could be related to each other.  For a second order phase transition the so called static critical exponents are found to be universal in a sense that apart from a few factors, like spatial dimensionality, symmetry of the system etc. they do not depend on the details of the interaction. As a matter of fact different physical systems may belong to the same universality class. Critical exponents are also found to satisfy certain \textit{scaling laws}, which in turn implies that not all the critical exponents are independent. These scaling relations are related to the \textit{scaling hypothesis} for thermodynamic functions \cite{stanley2}. 

Since black holes also behave as thermodynamic objects, therefore it would be natural to ask whether some or all the above features are present in the context of black hole mechanics.  An attempt along this direction had been commenced long back considering the charged and rotating black holes in asymptotically flat space time \cite{dav2}-\cite{hut}. Since then a number of investigations \cite{cr6}-\cite{newcr1} have been carried out in order to understand the scaling behavior of black holes in an asymptotically flat space time. In \cite{cr1}, the critical exponents for the Kerr-Newmann black holes were calculated for the first time to check the validity of the scaling laws for black holes near the critical point. Recently the thermodynamics of black holes in AdS space time has attained much attention in the context of AdS/CFT duality and  critical phenomena may play a crucial role in order to gain some insights regarding this duality. Like in the case of asymptotically flat space time, a number of attempts \cite{cr10}-\cite{cr7} have also been made in order to explore the critical behavior of rotating and charged black holes both in the de Sitter (dS) and anti de Sitter (AdS) space times. In spite of all these attempts, a detailed analysis of critical phenomena for black holes, based on a standard thermodynamic procedure, is still lacking in the literature. 

For the past two decades, Born-Infeld electrodynamics coupled to Einstein gravity has attracted renewed attention due to its several unique features \cite{prdw}-\cite{jhep}. A remarkable property of Born-Infeld electrodynamics \cite{born} is that it is able to describe a classical theory of charged particles with finite self energy while both the potential and electric field become regular at the origin. While eliminating the undesirable features of Maxwell electrodynamics it retains two of its very crucial aspects; namely, duality symmetry and causal physical propagation. These aspects taken together make the Born-Infeld like action a desirable candidate for a gravitational theory \cite{deser}. It is also to be noted that the Born-Infeld action naturally yields the higher derivative corrections to the usual Maxwell action, obtained by integrating out the matter fields in the gauge-matter sector. Thus even if there is no higher derivative correction to the Maxwell term from the beginning, such terms would arise due to loop effects from the matter sector. Moreover, in an appropriate limit, the standard Einstein-Maxwell theory is reproduced from the Einstein Born-Infeld action.

Now black hole solutions for Einstein gravity in AdS space coupled to Born-Infeld electrodynamics were found in \cite{fernandogrg} and can be realized as a generalization of the Reissner-Nordstrom AdS black holes. They were also shown \cite{fernando} to exhibit similar thermodynamic properties. Subsequently, although several attempts of the thermodynamic behavior of the asymptotically AdS black holes coupled to Born-Infeld electrodynamics were studied \cite{fernando}-\cite{olivera}, some basic issues remain unanswered. Specifically, a study of critical phenomena and universality classes is lacking. The motivation of our work is to fill this lacuna which would also explain the similarity of the thermodynamic behavior of Reissner-Nordstrom AdS black holes with their Born-Infeld generalization.  

In this paper, following a standard thermodynamic approach, we attempt to resolve several vexing issues for the specific example of Born-Infeld AdS (BI AdS) black holes. Based on Ehrenfest's scheme\cite{zeman}, appropriately tailored for black holes, we have successfully analyzed several thermodynamic aspects of phase transition phenomena in various black holes, in a series of papers \cite{bss}-\cite{dibakar2}. We now extend our formalism to examine critical phenomena for black holes with particular reference to the Born-Infeld AdS example. Considering a canonical ensemble we explicitly calculate the static critical exponents associated with the divergences of various thermodynamic entities (like the heat capacity at constant charge ($ C_{Q} $) etc.) near the critical point. All these exponents are found to satisfy the \textit{thermodynamic scaling laws} \cite{stanley1}. Furthermore we explore the Generalized Homogeneous Function (GHF) hypothesis \cite{stanley1} for the free energy near the critical point and show its compatibility with the thermodynamic scaling laws. Finally we check the validity of the additional scaling relations considering the spatial dimension ($ d $) of the system as three.  

Before we proceed further, let us briefly mention about the organization of our paper. In section 2 we calculate all the essential thermodynamic quantities that will be required to discuss the critical behavior of Born-Infeld AdS black holes. In section 3, based on a thermodynamic approach we systematically analyze the critical behavior of Born-Infeld AdS black holes. Finally we draw our conclusions in section 4.

\section{Thermodynamics of Born-Infeld black holes in AdS space}
The Einstein- Born-Infeld action in $ (3+1) $ dimensions is given by \cite{myung},
\begin{equation}
S= \int d^{4}x\sqrt{-g}\left[ \frac{R-2\Lambda}{16\pi G}+L(F)\right] 
\end{equation}
where,
\begin{equation}
L(F)=\frac{b^{2}}{4\pi G}\left( 1-\sqrt{1+\frac{2F}{b^{2}}}\right) 
\end{equation}
with $ F=\frac{1}{4} F_{\mu\nu}F^{\mu\nu} $. Here  $ b $ is the Born-Infeld parameter which is related to the string tension $ \alpha^{'} $ as $ b=1/2\pi \alpha^{'} $ and $ \Lambda(=-3/l^2) $ is the cosmological constant.   

By solving the equations of motion, the Born-Infeld anti de sitter (BI AdS) solution may be found as,
\begin{equation}
ds^2 = -\chi dt^2+\chi ^{-1}dr^2+r^2 d\Omega^{2}
\end{equation} 
where,
\begin{equation}
\chi(r) = 1-\frac{2M}{r}+\frac{r^{2}}{l^{2}}+\frac{2b^{2}r^{2}}{3}\left( 1-\sqrt{1+\frac{Q^{2}}{b^{2}r^{4}}}\right) 
+\frac{4Q^{2}}{3r^{2}} F\left( \frac{1}{4},\frac{1}{2},\frac{5}{4},-\frac{Q^{2}}{b^{2}r^{4}}\right)\label{chi},
\end{equation}
and $ F $ is a hypergeometric function \cite{hyper}. In the limit $ b\rightarrow\infty $ and $ Q\neq0 $ one obtains the corresponding solution for Reissner Nordstrom (RN) AdS black holes. Clearly this is a nonlinear generalization of the RN AdS black holes.

In order to obtain an expression for the ADM mass ($ M $) of the black hole we set $\chi(r_{+})=0$, which yields, $ (G=1) $
\begin{equation}
M= \frac{r_{+}}{2}+ \frac{r^{3}_{+}}{2l^{2}}+\frac{b^{2}r^{3}_{+}}{3}\left( 1-\sqrt{1+\frac{Q^{2}}{b^{2}r^{4}_{+}}}\right)
+\frac{2Q^{2}}{3r_{+}}\left( 1-\frac{Q^{2}}{10b^{2}r^{4}_{+}}\right)+ O(1/b^{4}) , 
\label{M}
\end{equation}
where $ r_{+} $ is the radius of the outer event horizon. Also, all the higher order terms from $ (1/b^{2})^{2} $ onwards have been dropped out from the series expansion of $ F\left( \frac{1}{4},\frac{1}{2},\frac{5}{4},-\frac{Q^{2}}{b^{2}r^{4}_{+}}\right) $. Following a similar spirit we can express the electrostatic potential difference ($ \Phi $) between the horizon and infinity as \cite{fernando},
\begin{equation}
\Phi=\frac{Q}{r_{+}}\left( 1-\frac{Q^{2}}{10b^{2}r^{4}_{+}}\right)+ O(1/b^{4}), 
\label{Phi}
\end{equation}
where $ Q $ is the electric charge. Henceforth all our results are valid upto $ O(1/b^{2}) $ only.

Using (\ref{chi}) and (\ref{M}), the Hawking temperature may be obtained as,
\begin{eqnarray}
T&=& \frac{\chi^{'}(r_{+})}{4\pi}\nonumber\\
&=&\frac{1}{4\pi}\left[ \frac{1}{r_{+}}+\frac{3r_{+}}{l^{2}}+2b^{2}r_{+}\left( 1-\sqrt{1+\frac{Q^{2}}{b^{2}r^{4}_{+}}}\right)  \right] \label{T}.
\end{eqnarray}

The entropy of the black hole is given by
\begin{eqnarray}
S=\int T^{-1}\left( \frac{\partial M}{\partial r_{+}}\right)_{Q} dr_{+}= \pi r^{2}_{+}\label{S}. 
\end{eqnarray}
As a next step we would like to compute the heat capacity at constant charge ($ C_{Q} $) in order to understand the critical behavior of Born-Infeld AdS black holes.
Using (\ref{T}) and (\ref{S}) the specific heat (at constant charge) may be found as
\begin{eqnarray}
C_{Q}=T\left(\frac{\partial S}{\partial T} \right)_{Q}  &=& T \frac{\left( \partial S/\partial r_{+}\right)_{Q}}{\left( \partial T/\partial r_{+}\right)_{Q}}\nonumber\\
&=&\frac{2\pi r^{2}_{+}\sqrt{1+\frac{Q^{2}}{b^{2}r^{4}_{+}}}\left[r^{2}_{+}+\lambda r^{4}_{+}+2b^{2}r^{4}_{+}\left( 1-\sqrt{1+\frac{Q^{2}}{b^{2}r^{4}_{+}}}\right) \right]}{r^{2}_{+}(\lambda r^{2}_{+}-1)\sqrt{1+\frac{Q^{2}}{b^{2}r^{4}_{+}}}-2b^{2}r^{4}_{+}\left( 1-\sqrt{1+\frac{Q^{2}}{b^{2}r^{4}_{+}}}\right)+2Q^{2}}\label{CQ},
\end{eqnarray}
where $ \lambda=-\Lambda=3/l^{2} $.

From (\ref{CQ}) we note that in order to have a divergence in $ C_Q $ one must satisfy the following condition, 
\begin{equation}
r^{2}_{+}(\lambda r^{2}_{+}-1)\sqrt{1+\frac{Q^{2}}{b^{2}r^{4}_{+}}}-2b^{2}r^{4}_{+}\left( 1-\sqrt{1+\frac{Q^{2}}{b^{2}r^{4}_{+}}}\right)+2Q^{2}=0.\label{root}
\end{equation}
It is problematic to provide an analytical solution for an arbitrary choice of parameters. Nevertheless it is possible to give certain plausibility arguments that give a condition on the parameters to yield positive roots for $ r_+ $. Let us take the extremal BI AdS black hole. Here both $ \chi(r) $ and $ \frac{d\chi}{dr} $ vanish at the degenerate horizon ($ r_e $). The corresponding equation is now given by \cite{myung} 
\begin{equation}
1+ \left(2b^{2}+\frac{3}{l^{2}} \right)r^{2}_{e} -2b^{2}\sqrt{r^{4}_{e}+\frac{Q^{2}}{b^{2}}}=0
\end{equation}   
whose solution is,
\begin{equation}
r^{2}_{e}=\frac{l^{2}}{6}\left( \frac{1+\frac{3}{2b^{2}l^{2}}}{1+\frac{1}{b^{2}l^{2}}}\right) \left[ -1+\sqrt{1+\frac{12(1+\frac{3}{4b^{2}l^{2}})}{b^{2}l^{2}(1+\frac{3}{4b^{2}l^{2}})^{2}}\left(b^{2}Q^{2}-\frac{1}{4} \right)} \right]. 
\end{equation}
In order to have a real root we must have $ bQ \geq 0.5 $. The forbidden region for BI AdS black holes is thus given by $ 0\leq bQ <0.5 $ while the allowed region is,
\begin{equation}
0.5\leq bQ\leq\infty.\label{ext}
\end{equation}
The special case $ b\rightarrow\infty $ just corresponds to the RN AdS black hole.

Requiring that a smooth extremal limit holds, we choose the parameter space satisfying the condition (\ref{ext}) in order to find out the roots of the equation (\ref{root}). Numerical analysis reveals that now there are two positive and two negative roots for $ r_+ $. We therefore pursue our analysis subject to the
condition (\ref{ext}).

In tables ($ 1a $), (\ref{E2}) and (\ref{E3}) we give the solution for (\ref{root}) obtained by numerical methods subject to the condition (\ref{ext}). The value $ bQ=0.5 $ when the bound is saturated is also considered (last entry in table ($ 1a $)). From the numerical solutions it is quite evident that, for a general choice of parameters ($ Q $ and $ b $), equation (\ref{root}) possesses four distinct real roots implying that the denominator of (\ref{CQ}) has only simple poles. Among these four roots there are two positive and two negative roots. The phase transition occurs at the two positive roots ($ r_1 $ and $ r_2 $). This fact has also been depicted in various figures (1, 2, 3, 4, 5, 6, 7 and 8) for different sets of parameters. For the specific choice of $ Q=0.5 $, $ b=10 $ (and $ l=10 $), which is the example considered in details, the roots of the equation (\ref{root}) are $ \pm 0.875061$ and $\pm 5.70663$.  We take only the two positive roots ($ r_1=0.875061 $, $ r_2=5.70663 $) which are also marked on the various figures (1 and 2).

\begin{subtables}
\begin{table}[htb]
\caption{Roots of equation (\ref{root}) for $ b=10 $ and $ l=10 $}   
\centering                          
\begin{tabular}{c c c c c c c}            
\hline\hline                        
$Q$ & $r_1$  & $r_2$ &$r_3$ & $r_4 $ &  \\ [0.05ex]
\hline
0.5 & 0.875061 & 5.70663 & -0.875061 & -5.70663 \\
0.4 & 0.696540 & 5.73116 & -0.696540 & -5.73116 \\                              
0.3 & 0.519874 & 5.74988 & -0.519874 & -5.74988 \\
0.2 & 0.344194 & 5.76306 & -0.344194 & -5.76306 \\ 
0.1 & 0.167309 & 5.77090 & -0.167309 & -5.77090 \\ 
0.05& 0.0707235& 5.77285 & -0.0707235& -5.77285 \\ [0.05ex]         
\hline                              
\end{tabular}\label{E1}  
\end{table}
\begin{table}[htb]
\caption{Roots of equation (\ref{root}) for $ b=15 $ and $ l=10 $}   
\centering                          
\begin{tabular}{c c c c c c c}            
\hline\hline                        
$Q$ & $r_1$  & $r_2$ &$r_3$ & $r_4 $ &  \\ [0.05ex]
\hline
0.5 & 0.875680 & 5.70663 & -0.875680 & -5.70663 \\
0.4 & 0.697319 & 5.73116 & -0.697319 & -5.73116 \\                              
0.3 & 0.520919 & 5.74988 & -0.520919 & -5.74988 \\
0.2 & 0.345784 & 5.76306 & -0.345784 & -5.76306 \\ 
0.1 & 0.170722 & 5.77090 & -0.170722 & -5.77090 \\ [0.05ex]         
\hline                              
\end{tabular}
\label{E2}          
\end{table}
\begin{table}[htb]
\caption{Roots of equation (\ref{root}) for $ b=20 $ and $ l=10 $}   
\centering                          
\begin{tabular}{c c c c c c c}            
\hline\hline                        
$Q$ & $r_1$  & $r_2$ &$r_3$ & $r_4 $ &  \\ [0.05ex]
\hline
0.5 & 0.875896 & 5.70663 & -0.875896 & -5.70663 \\
0.4 & 0.697590 & 5.73116 & -0.697590 & -5.73116 \\                              
0.3 & 0.521283 & 5.74988 & -0.521283 & -5.74988 \\
0.2 & 0.346335 & 5.76306 & -0.346335 & -5.76306 \\ 
0.1 & 0.171859 & 5.77090 & -0.171859 & -5.77090 \\ [0.05ex]         
\hline                              
\end{tabular}
\label{E3}          
\end{table}
\end{subtables}

\begin{figure}[h]
\centering
\includegraphics[angle=0,width=8cm,keepaspectratio]{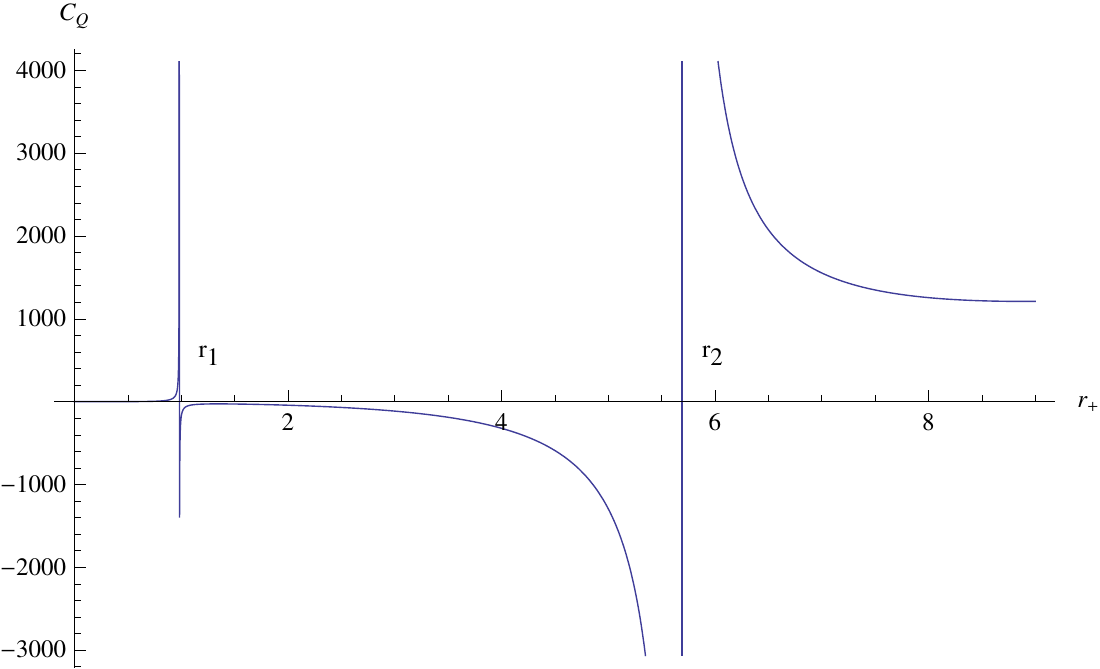}
\caption[]{\it Specific heat plot ($ C_{Q} $) for Born-Infeld AdS black hole with respect to $r_{+}$ for $Q(=Q_c)=0.5$, $ b=10 $ and $ l=10 $. The discontinuity at $ r_2 $ is shown while that at $ r_1 $ is more clearly shown in figure 2.}
\label{figure 2a}
\end{figure} 
\begin{figure}[h]
\centering
\includegraphics[angle=0,width=8cm,keepaspectratio]{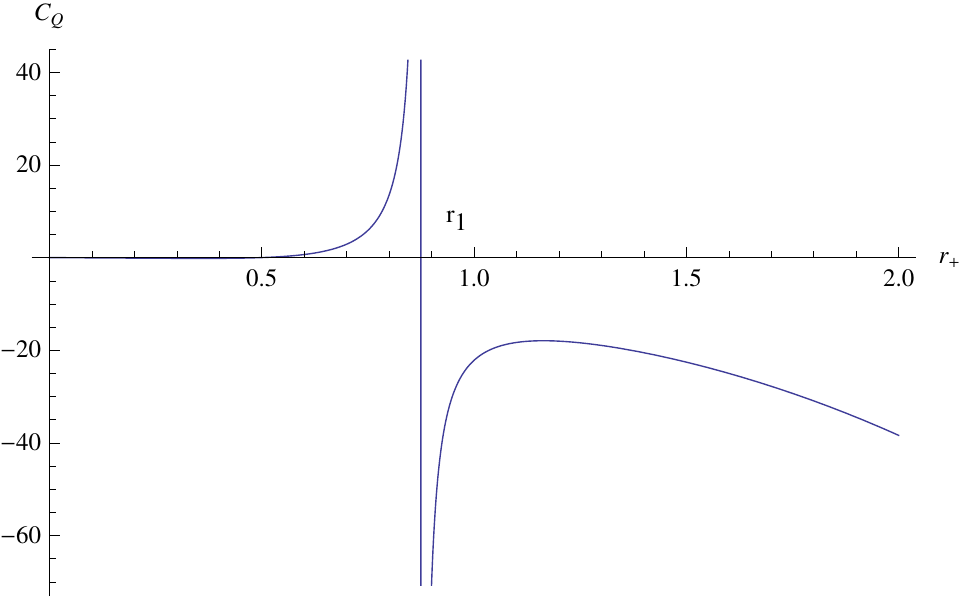}
\caption[]{\it Discontinuity of Specific heat ($ C_{Q} $) for Born-Infeld AdS black hole at $r_{+}= r_1$ for $Q(=Q_c)=0.5$, $ b=10 $ and $ l=10 $}
\label{figure 2a}
\end{figure} 
\begin{figure}[h]
\centering
\includegraphics[angle=0,width=8cm,keepaspectratio]{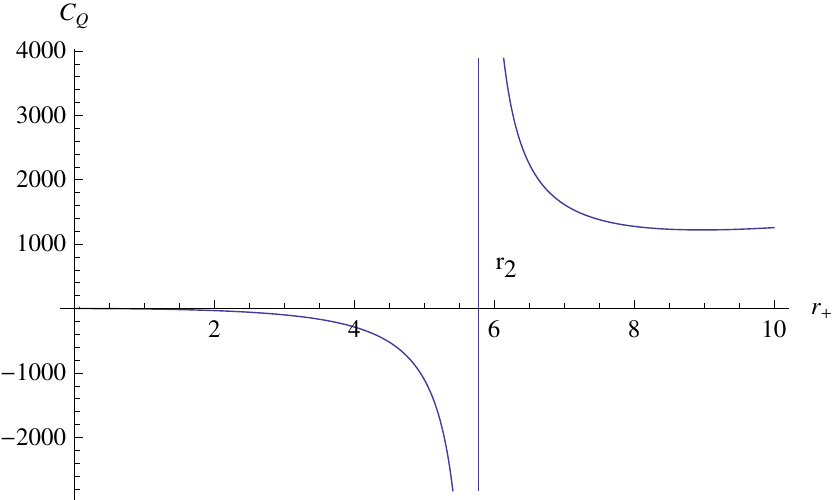}
\caption[]{\it Discontinuity of Specific heat ($ C_{Q} $) for Born-Infeld AdS black hole at $r_{+}= r_2$ for $Q(=Q_c)=0.05$, $ b=10 $ and $ l=10 $}
\label{figure 2a}
\end{figure}
\begin{figure}[h]
\centering
\includegraphics[angle=0,width=8cm,keepaspectratio]{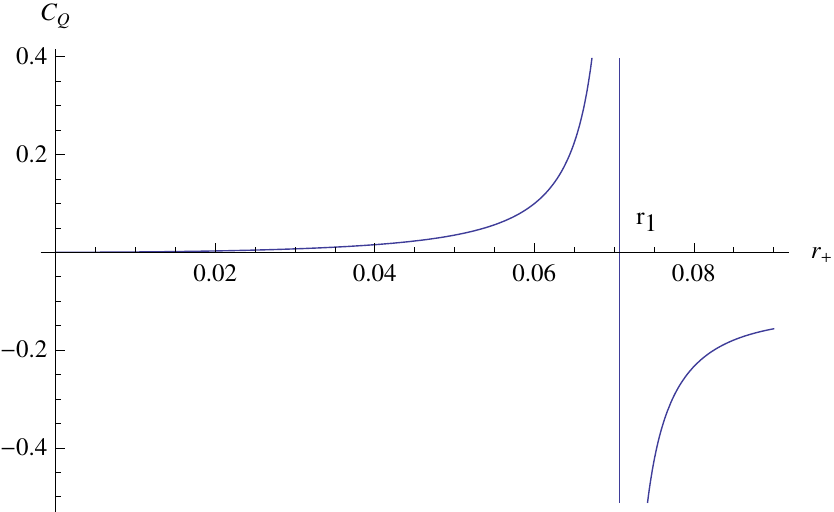}
\caption[]{\it Discontinuity of Specific heat ($ C_{Q} $) for Born-Infeld AdS black hole at $r_{+}= r_1$ for $Q(=Q_c)=0.05$, $ b=10 $ and $ l=10 $}
\label{figure 2a}
\end{figure} \begin{figure}[h]
\centering
\includegraphics[angle=0,width=8cm,keepaspectratio]{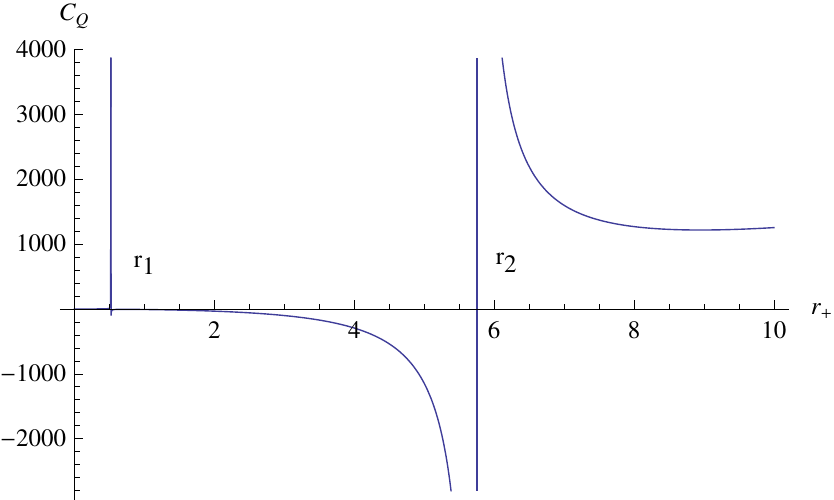}
\caption[]{\it Discontinuity of Specific heat ($ C_{Q} $) for Born-Infeld AdS black hole at $r_{+}= r_2$ for $Q(=Q_c)=0.3$, $ b=15 $ and $ l=10 $}
\label{figure 2a}
\end{figure}
\begin{figure}[h]
\centering
\includegraphics[angle=0,width=8cm,keepaspectratio]{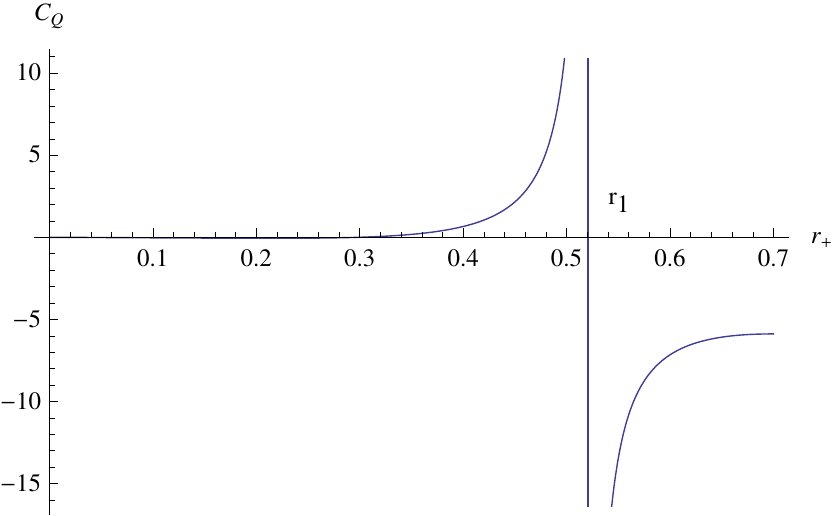}
\caption[]{\it Discontinuity of Specific heat ($ C_{Q} $) for Born-Infeld AdS black hole at $r_{+}= r_1$ for $Q(=Q_c)=0.3$, $ b=15 $ and $ l=10 $}
\label{figure 2a}
\end{figure} 
\begin{figure}[h]
\centering
\includegraphics[angle=0,width=8cm,keepaspectratio]{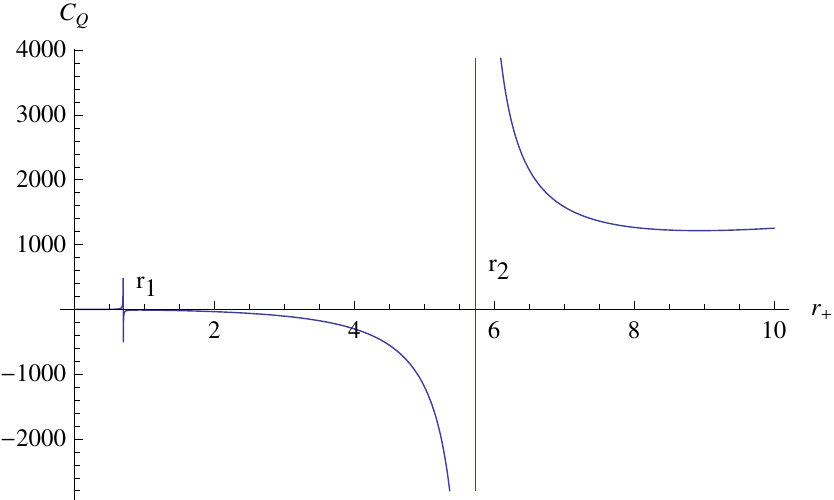}
\caption[]{\it Discontinuity of Specific heat ($ C_{Q} $) for Born-Infeld AdS black hole at $r_{+}= r_2$ for $Q(=Q_c)=0.4$, $ b=20 $ and $ l=10 $}
\label{figure 2a}
\end{figure} 
\begin{figure}[h]
\centering
\includegraphics[angle=0,width=8cm,keepaspectratio]{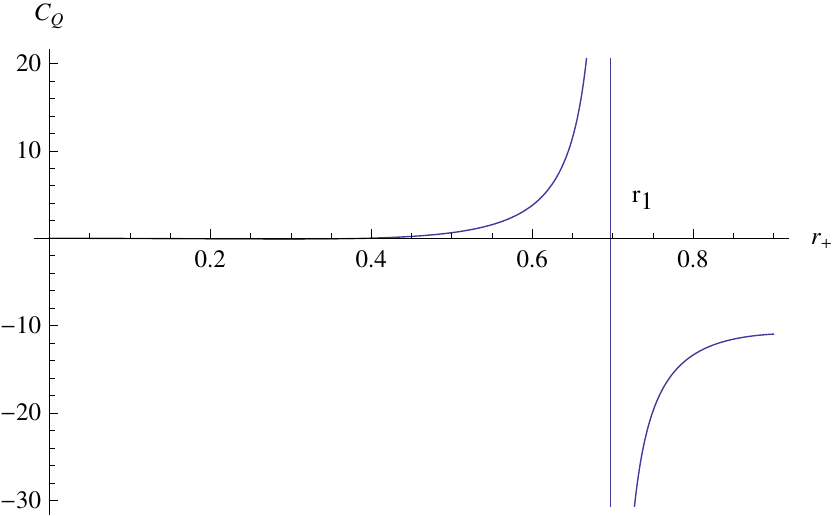}
\caption[]{\it Discontinuity of Specific heat ($ C_{Q} $) for Born-Infeld AdS black hole at $r_{+}= r_1$ for $Q(=Q_c)=0.4$, $ b=20 $ and $ l=10 $}
\label{figure 2a}
\end{figure}   

From figure 1 and figure 2 we observe that $ C_{Q} $ suffers discontinuities at two points, namely $ r_{1} $ and $ r_{2} $(discussed in the previous paragraph), which may be identified as the critical points for the phase transition phenomena in BI AdS black holes. From the above figures it is evident that the heat capacity is positive for $ r_+ < r_1 $ and $ r_+ >r_2 $, while it is negative in the intermediate range $ r_1<r_+<r_2 $.  Since the  black hole with smaller mass possesses lesser entropy/horizon radius than the black hole with larger mass, therefore the point $ r_{1} $ corresponds to the critical point for the transition between a smaller mass black hole with positive specific heat ($ C_Q>0 $) to an intermediate unstable black hole with negative heat capacity ($ C_Q<0 $). On the other hand  $ r_{2} $ corresponds to a transition from the intermediate unstable black hole to a larger mass black hole with positive heat capacity ($ C_Q>0 $) \cite{myung}.  

In order to calculate the isothermal compressibility related derivative $ K^{-1}_{T} $ we first note that,
\begin{eqnarray}
 K^{-1}_{T}=Q\left( \partial\Phi /\partial Q\right)_{T} =- Q \left(\frac{\partial \Phi}{\partial T} \right)_{Q} \left(\frac{\partial T}{\partial Q} \right)_{\Phi},
\end{eqnarray}
where we have used the thermodynamic identity $\left(\frac{\partial \Phi}{\partial T} \right)_{Q}\left(\frac{\partial T}{\partial Q} \right)_{\Phi}\left(\frac{\partial Q}{\partial \Phi} \right)_{T} =-1$.

Finally, using (\ref{Phi}) and (\ref{T}) the expression for $ K^{-1}_{T} $ may be found as,
\begin{equation}
K^{-1}_{T}=\frac{\wp(Q,r_+)}{\Re(Q,r_+)}\label{kt}
\end{equation}
where,
\begin{eqnarray}
\wp(Q,r_+)= \left( \frac{Q}{r_{+}}\right) \left( 1-\frac{3Q^{2}}{10b^{2}r^{4}_{+}}\right)\left[ r^{2}_{+}(\lambda r^{2}_{+}-1)\sqrt{1+\frac{Q^{2}}{b^{2}r^{4}_{+}}}-2b^{2}r^{4}_{+}\left( 1-\sqrt{1+\frac{Q^{2}}{b^{2}r^{4}_{+}}}\right)+2Q^{2}\right]\nonumber\\
-\frac{2Q^{3}}{r_{+}}\left( 1-\frac{Q^{2}}{2b^{2}r^{4}_{+}}\right)
\end{eqnarray}
and
\begin{equation}
\Re(Q,r_+) = r^{2}_{+}(\lambda r^{2}_{+}-1)\sqrt{1+\frac{Q^{2}}{b^{2}r^{4}_{+}}}-2b^{2}r^{4}_{+}\left( 1-\sqrt{1+\frac{Q^{2}}{b^{2}r^{4}_{+}}}\right)+2Q^{2}.
\end{equation}
From the above expressions one finds that $ K^{-1}_{T} $ diverges exactly at the points where the heat capacity ($ C_{Q} $) diverges. A similar conclusion holds for the Kerr Newmann black hole in asymptotically flat space \cite{cr4} and is compatible with general thermodynamic arguments\cite{dt}. With all these relevant expressions in hand, we are now in a position to investigate the critical behavior of BI AdS black holes near the critical points $ r_{1} $ and $ r_{2} $.

Before concluding the section we remark that similar results also hold for the other sets of parameters considered in figures (3,~4,~5,~6,~7~ and~ 8).

\section{Critical exponents and scaling laws}
In the previous section, through numerical analysis, we have found that equation (\ref{root}) possesses two distinct positive roots ($ r_1 $ and $ r_2 $) (subjected to the condition (\ref{ext})) which are also the critical points for the phase transition occurring in BI AdS black holes. In this section we aim to perform an analytical investigation of the critical behavior of BI AdS black holes associated with these critical points.  

The critical point is marked a divergence in the heat capacity. It is important to understand the nature of this divergence and the singular behavior of other thermodynamic functions near the critical point. In order to do this we introduce a set of critical exponents $ (\alpha,\beta,\gamma,\delta,\varphi,\psi,\nu,\eta) $ which play a central role in the theory of critical phenomena. These critical exponents are associated with the discontinuities of various thermodynamical variables. They are to a large degree universal, depending only on a few fundamental parameters like the dimensionality of the space, symmetry of the order parameter etc..   
\begin{figure}[h]
\centering
\includegraphics[angle=0,width=7cm,keepaspectratio]{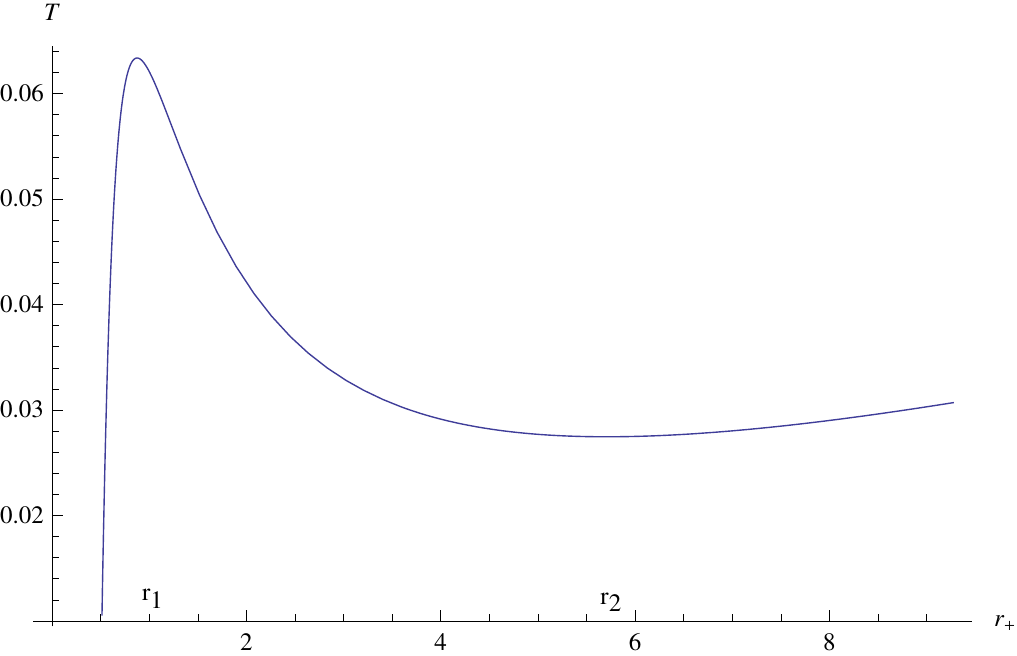}
\caption[]{\it Hawking temperature plot ($ T $) for Born-Infeld AdS black hole with respect to $r_{+}$ for $Q(=Q_c)=0.5$, $ b=10 $ and $ l=10 $}
\label{figure 2a}
\end{figure} 

Let us first calculate the critical exponent ($ \alpha $) that is associated with the divergences of the heat capacity ($ C_{Q} $) near the critical points. In order to do that let us first note that near the critical points ($ r_i $) we can write
\begin{equation}
r_{+}=r_{i}(1+\Delta), ~~~~ i=1,2 \label{tri}
\end{equation}
where $|\Delta| <<1 $.
As already discussed there are two distinct positive roots for the critical point $ r_i $ ($ r_1 $ and $ r_2 $). Also, any function of $ r_+ $, in particular the temperature $ T(r_+) $, may be expressed as
\begin{equation}
T(r_+)=T(r_i)(1+\epsilon)
\end{equation}
where $ |\epsilon|<<1$.
As a next step, for a fixed value of the charge ($ Q $), we Taylor expand $ T(r_+) $ in a sufficiently small neighborhood of $ r_i $ which yields,
\begin{eqnarray}
T(r_+)= T(r_i)+\left[ \left( \frac{\partial T}{\partial r_+}\right)_{Q=Q_c}\right]_{r_+=r_i} (r_+-r_i)+\frac{1}{2} \left[ \left( \frac{\partial^{2} T}{\partial r^{2}_+}\right)_{Q=Q_c}\right]_{r_+=r_i} (r_+-r_i)^{2}\nonumber\\
+ higher~~ order~~ terms.\label{Tr}
\end{eqnarray}
Since $ C_Q $ diverges at $ r_+=r_i $, therefore the second term on the R.H.S. of (\ref{Tr}) vanishes by virtue of equation (\ref{CQ}).  
Using (\ref{tri}) we finally obtain from (\ref{Tr})
\begin{equation}
\Delta=\frac{\epsilon^{1/2}}{D_{i}^{1/2}}\label{delta}
\end{equation} 
where\footnote{We use the notation $ T(r_i)=T_i $.},
\begin{eqnarray}
D_i &=& \frac{r^{2}_{i}}{2T_{i}}\left[ \left( \frac{\partial^{2} T}{\partial r^{2}_+}\right)_{Q=Q_c}\right]_{r_+=r_i}\nonumber\\
&=& \frac{r^{2}_{i}\left(1+\frac{Q^{2}_{c}}{b^{2}r^{4}_{i}}\right)^{3/2}-6Q^{2}_{c}\left(1+\frac{Q^{2}_{c}}{b^{2}r^{4}_{i}}\right)+\frac{4Q^{4}_{c}}{b^{2}r^{4}_{i}}}{4\pi r^{3}_{i}T_i \left(1+\frac{Q^{2}_{c}}{b^{2}r^{4}_{i}}\right)^{3/2}}.
\end{eqnarray}
From figure 9 it is to be noted that in the neighborhood of $ r_+=r_2 $ we always have $ T(r_+)>T(r_2) $ so that $ \epsilon $ is positive. On the other hand for any point close to  $ r_+=r_1 $ we have $ T(r_+)<T(r_1) $ implying that $ \epsilon $ is negative. Therefore, based on this observation and substituting $ r_+ $ from (\ref{tri}) into (\ref{CQ}) and using (\ref{delta}) the singular behavior of $ C_Q $ near the critical point $ r_+=r_2 $ may be found as,
\begin{eqnarray}
C_Q \simeq ~~~\left[ \frac{A_{i}}{\epsilon^{1/2}}\right] _{r_i=r_2}
\end{eqnarray}
where,
\begin{equation}
A_i=\frac{\pi r^{2}_{i}D^{1/2}_{i}\sqrt{1+\frac{Q^{2}_{c}}{b^{2}r^{4}_{i}}}\left[r^{2}_{i}+\lambda r^{4}_{i}+2b^{2}r^{4}_{i}\left( 1-\sqrt{1+\frac{Q^{2}_{c}}{b^{2}r^{4}_{i}}}\right) \right]}{r^{2}_{i}(2\lambda r^{2}_{i}-1)\sqrt{1+\frac{Q^{2}_{c}}{b^{2}r^{4}_{i}}}-\frac{Q^{2}_{c}}{b^{2}r^{2}_{i}}(\lambda r^{2}_{i}-1)}.
\end{equation}
Note that in the above expression we have retained terms only linear in $ \Delta $ while expanding the denominator of $ C_Q $ near the critical point. On the other hand, following a similar approach, the singular behavior of $ C_Q $ near $ r_+=r_1 $ may be expressed as,
\begin{equation}
C_Q \simeq  ~~~\left[ \frac{A_{i}}{(-\epsilon)^{1/2}}\right] _{r_i=r_1}.
\end{equation}
Combining both of these facts into a single expression, we may therefore express the singular behavior of the heat capacity ($ C_Q $) near the critical points as,
\begin{eqnarray}
C_Q &\simeq & ~~~ \frac{A_{i}}{|\epsilon|^{1/2}}\nonumber\\
&=&~~~\frac{A_{i}T^{1/2}_{i}}{|T-T_i|^{1/2}}\label{CQ1}.
\end{eqnarray}

Comparing (\ref{CQ1}) with the standard form
\begin{equation}
C_Q \sim |T-T_i|^{-\alpha}
\end{equation}
we find $ \alpha=1/2 $.

Next, we want to calculate the critical exponent $ \beta $ which is related to the electric potential ($ \Phi $) for a fixed value of charge as,
\begin{equation}
\Phi(r_+) - \Phi(r_i) \sim |T-T_i|^{\beta}\label{Ph}.
\end{equation}
In order to do that we Taylor expand $ \Phi(r_+) $ close to the critical point $ r_+=r_i $ which yields, 
\begin{eqnarray}
\Phi(r_+)= \Phi(r_i)+\left[ \left( \frac{\partial \Phi}{\partial r_+}\right)_{Q=Q_c}\right]_{r_+=r_i} (r_+-r_i)
+ higher~~ order~~ terms.\label{phir}
\end{eqnarray}
Ignoring all the higher order terms in (\ref{phir}) and using (\ref{Phi}) and (\ref{delta}) we finally obtain
\begin{equation}
\Phi(r_+)- \Phi(r_i)= - \left( \frac{Q_c}{r_i T^{1/2}_i D^{1/2}_{i}}\right) \left(1-\frac{Q^{2}_{c}}{2b^{2}r^{4}_{i}} \right) |T-T_i|^{1/2}\label{Pr}.
\end{equation} 
Comparing (\ref{Ph}) and (\ref{Pr}) we find $ \beta=1/2 $.

Let us now calculate the critical exponent $ \gamma $ which is related to the singular behavior of $ K^{-1}_{T} $ (near the critical points $ r_i $) for a fixed value of charge ($ Q=Q_c $) as \cite{cr4},\cite{cr13}
\begin{equation}
K^{-1}_{T}\sim |T-T_i|^{-\gamma}\label{k}.
\end{equation}
Following our previous approach, we substitute $ r_+ $ from (\ref{tri}) into (\ref{kt}) and use (\ref{delta}) which finally yields,
\begin{equation}
K^{-1}_{T}=\frac{B_i}{|\epsilon|^{1/2}}= \frac{B_i T^{1/2}_i}{|T-T_i|^{1/2}}\label{KT}
\end{equation}   
where,
\begin{equation}
B_i=\frac{D^{1/2}_i\wp(Q_c,r_i)}{2r^{2}_{i}(2\lambda r^{2}_{i}-1)\sqrt{1+\frac{Q^{2}_{c}}{b^{2}r^{4}_{i}}}-\frac{2Q^{2}_{c}}{b^{2}r^{2}_{i}}(\lambda r^{2}_{i}-1)}.
\end{equation}
From (\ref{k}) and (\ref{KT}) we note that $ \gamma=1/2 $.
\begin{figure}[h]
\centering
\includegraphics[angle=0,width=7cm,keepaspectratio]{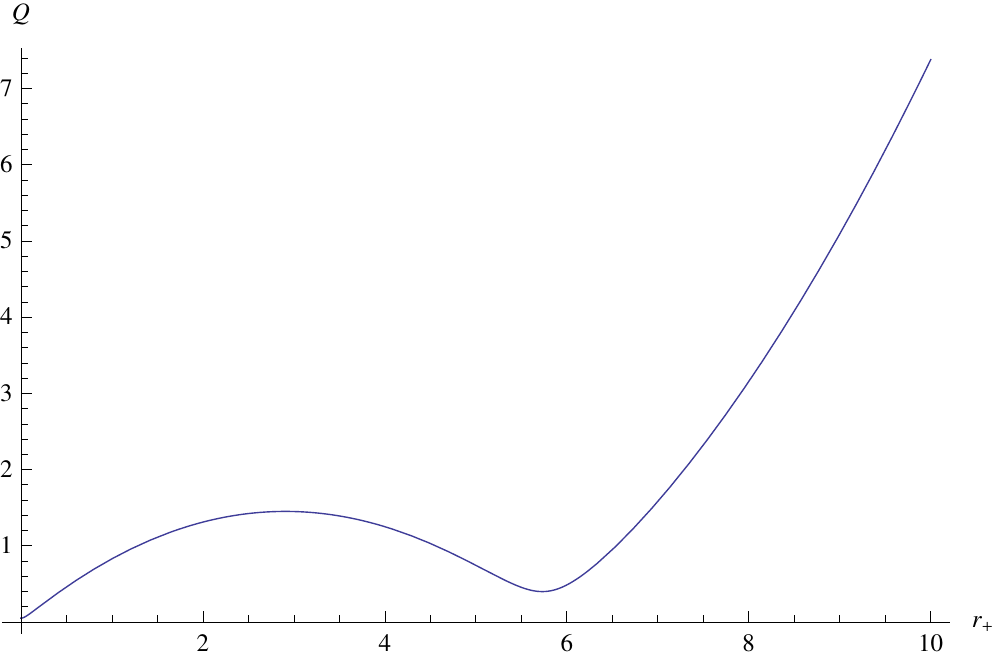}
\caption[]{\it Charge ($ Q $) for Born-Infeld AdS black hole with respect to $r_{+}$ for $T(=T_2)=0.0275$, $ b=10 $ and $ l=10 $}
\label{figure 2a}
\end{figure} 

Let us now calculate the critical exponent ($ \delta $) which is related to the electric potential ($ \Phi $) for the fixed value of the temperature $ T=T_i $ as,
\begin{equation}
\Phi(r_+) - \Phi(r_i) \sim |Q-Q_i|^{1/\delta}\label{PQ},
\end{equation}
where $ Q_i $ is the value of charge ($ Q $) at $ r_+=r_i $. 
In order to do that we first expand $ Q(r_+) $ in a sufficiently small neighborhood of $r_+= r_i $ which yields,
 \begin{eqnarray}
Q(r_+)= Q(r_i)+\left[ \left( \frac{\partial Q}{\partial r_+}\right)_{T=T_i}\right]_{r_+=r_i} (r_+-r_i)+\frac{1}{2} \left[ \left( \frac{\partial^{2} Q}{\partial r^{2}_+}\right)_{T=T_i}\right]_{r_+=r_i} (r_+-r_i)^{2}\nonumber\\
+ higher~~ order~~ terms.\label{Q}
\end{eqnarray}
Using the functional form
\begin{equation}
T=T(r_+,Q)
\end{equation}
and following our previous argument we note that,
\begin{equation}
\left[ \left( \frac{\partial Q}{\partial r_+}\right)_{T}\right]_{r_+=r_i}=-\left[ \left( \frac{\partial T}{\partial r_+}\right)_{Q}\right]_{r_+=r_i} \left( \frac{\partial Q}{\partial T}\right)_{r_+=r_i}=0 \label{e2}.
\end{equation}
Also note that near the critical point we can express the charge ($ Q $) as,
\begin{equation}
Q(r_+)= Q(r_i)(1+\Pi) \label{Qr}
\end{equation}
with $ |\Pi|<<1 $.  
Finally using (\ref{tri}) and (\ref{Qr}) from (\ref{Q}) we obtain 
\begin{equation}
\Delta=\left(\frac{2Q_i}{M_i r^{2}_{i}} \right)^{1/2}\Pi^{1/2}\label{e4}
\end{equation}  
where,
\begin{eqnarray}
M_i&=&\left[ \left( \frac{\partial^{2} Q}{\partial r^{2}_+}\right)_{T}\right]_{r_+=r_i}\nonumber\\
&=&\frac{r^{2}_{i}\left( 1+\frac{Q^{2}_{c}}{b^{2}r^{4}_{i}}\right) (3\lambda r^{2}_{i}+6b^{2}r^{2}_i-1)-\sqrt{1+\frac{Q^{2}_{c}}{b^{2}r^{4}_{i}}}(6b^{2}r^{4}_{i}+2Q^{2}_c)-\frac{2Q^{2}_c}{b^{2}r^{2}_i}(\lambda r^{2}_i-1)-4Q^{2}_c}{2Q_cr^{2}_i \sqrt{1+\frac{Q^{2}_{c}}{b^{2}r^{4}_{i}}}} \label{Mi}
\end{eqnarray}
Let us now consider the functional relation
\begin{equation}
\Phi=\Phi(r_+,Q)
\end{equation}
from which we find,
\begin{equation}
\left[ \left( \frac{\partial \Phi}{\partial r_+}\right)_{T} \right]_{r_+=r_i} = \left[ \left( \frac{\partial \Phi}{\partial r_+}\right)_{Q} \right]_{r_+=r_i}+\left[ \left( \frac{\partial Q}{\partial r_+}\right)_{T}\right]_{r_+=r_i}\left( \frac{\partial \Phi}{\partial Q}\right)_{r_+=r_i}\label{e1}.
\end{equation}
Once again the second term on the R.H.S. of (\ref{e1}) vanishes by virtue of (\ref{e2}). Therefore by using (\ref{Phi}) we finally obtain
\begin{equation}
\left[ \left( \frac{\partial \Phi}{\partial r_+}\right)_{T} \right]_{r_+=r_i} =-\frac{Q_c}{r^{2}_i}\left( 1-\frac{Q^{2}_c}{2b^{2}r^{4}_i}\right)\label{e5} .
\end{equation}  
As a next step, for the fixed value of the temperature ($ T $) we Taylor expand $ \Phi(r_+,T) $ close to the critical point $ r_+=r_i $ which yields,
\begin{eqnarray}
\Phi(r_+)= \Phi(r_i)+\left[ \left( \frac{\partial \Phi}{\partial r_+}\right)_{T=T_i}\right]_{r_+=r_i} (r_+-r_i)
+ higher~~ order~~ terms\label{e3}.
\end{eqnarray}
Ignoring all the higher order terms in (\ref{e3}) and using (\ref{e4}) and (\ref{e5}) we finally obtain 
\begin{equation}
\Phi(r_+)-\Phi(r_i)=-\left( \frac{2}{M_i}\right) ^{1/2}\left[ \frac{Q_c}{r^{2}_i}\left( 1-\frac{Q^{2}_c}{2b^{2}r^{4}_i}\right)\right] |Q-Q_i|^{1/2}\label{e6}.
\end{equation} 
Comparing (\ref{e6}) with (\ref{PQ}) we note that $ \delta=2 $.

In order to calculate the critical exponent $ \varphi $ we note from (\ref{CQ1}) that near the critical point $ r_+ =r_i$ the heat capacity behaves as
\begin{equation}
C_Q \sim \frac{1}{\Delta}.
\end{equation}
Finally using (\ref{e4}) we note that near the critical point,
\begin{equation}
C_Q \sim \frac{1}{|Q-Q_i|^{1/2}} \label{e10}.
\end{equation}
Comparing (\ref{e10}) with the standard relation
\begin{equation}
C_Q \sim \frac{1}{|Q-Q_i|^{\varphi}}.
\end{equation}
we note that $ \varphi=1/2 $.

In an attempt to calculate the critical exponent $ \psi $,  we use (\ref{S}) and (\ref{tri}) in order to expand the entropy ($ S $) near the critical point $ r_+=r_i $. Then it follows, 
\begin{eqnarray}
S(r_+)= S(r_i)+ 2\pi r^{2}_i \Delta \label{e7}
\end{eqnarray}
where we have ignored the higher order terms in $ \Delta $ as we did earlier.
Finally using (\ref{e4}) we obtain,
\begin{equation}
S(r_+)-S(r_i)=\frac{2^{3/2}\pi r_i}{M^{1/2}_i}|Q-Q_i|^{1/2}\label{e8}.
\end{equation}
Comparing (\ref{e8}) with the standard form
\begin{equation}
S(r_+)-S(r_i) \sim |Q-Q_i|^{\psi}
\end{equation}
we note that $ \psi=1/2 $.

Before going into our subsequent discussions let us first tabulate the critical exponents.
\begin{table}[h]
\caption{Various critical exponents and their values}   
\centering                          
\begin{tabular}{c c c c c c c c c}            
\hline\hline                        
$Critical~Exponents  $ & $\alpha$ & $\beta$  & $\gamma$ & $ \delta $ & $\varphi $ & $ \psi $ \\ [2.0ex]
\hline
Values & 1/2 & 1/2 & 1/2 & 2 & 1/2 & 1/2\\ [2.0ex]         
\hline                              
\end{tabular}
\label{E1}          
\end{table}

In usual thermodynamics various critical exponents are found to satisfy certain (scaling) relations among themselves, known as \textit{thermodynamic scaling laws} \cite{stanley1}, which may be expressed as,
\begin{eqnarray}
\alpha + 2\beta +\gamma = 2,~~~\alpha +\beta(\delta + 1) =2,~~~(2-\alpha)(\delta \psi -1)+1 =(1-\alpha)\delta ~~\nonumber\\
~~ \gamma(\delta +1) =(2-\alpha)(\delta -1),~~~\gamma =\beta (\delta - 1),~~~\varphi + 2\psi -\delta^{-1} =1 \label{scaling laws}.
\end{eqnarray}     
It is interesting to note that, in the present context, all the above relations (\ref{scaling laws}) are indeed satisfied for BI AdS black holes.

We are now in a position to explore the Generalized Homogeneous Function (GHF) hypothesis \cite{cr1},\cite{cr8},\cite{stanley1} for the BI AdS black holes which states that \textit{close to the critical point, the singular part of the Helmholtz free energy $ F(T,Q)= M-TS $ is a generalized homogeneous function of its variables} i.e; there exists two numbers(known as scaling parameters) $ p $ and $ q $ such that for all positive $ \zeta $
\begin{equation}
F(\zeta^{p}\epsilon,\zeta^{q}\Pi)=\zeta F(\epsilon,\Pi)\label{ghf}.
\end{equation}
Let us first Taylor Expand $ F(T,Q) $ close to the critical point $ r_+=r_i $ which yields,
\begin{eqnarray}
F(T,Q)= F(T,Q)|_{r_+=r_i} + \left[ \left( \frac{\partial F}{\partial T}\right)_{Q}\right]_{r_+=r_i} (T-T_i)+\left[ \left( \frac{\partial F}{\partial Q}\right)_{T}\right]_{r_+=r_i} (Q-Q_i)\nonumber\\
+\frac{1}{2} \left[ \left( \frac{\partial^{2} F}{\partial T^{2}}\right)_{Q}\right]_{r_+=r_i} (T-T_i)^{2}+\frac{1}{2} \left[ \left( \frac{\partial^{2} F}{\partial Q^{2}}\right)_{T}\right]_{r_+=r_i} (Q-Q_i)^{2}+ other ~~~ terms \label{F1}
\end{eqnarray}
Since, 
\begin{equation}
C_Q= - T \left( \frac{\partial^{2} F}{\partial T^{2}}\right)_{Q} ~~~ and ~~~ K^{-1}_{T}= Q \left( \frac{\partial^{2} F}{\partial Q^{2}}\right)_{T}
\end{equation}
diverge at the critical point $ r_+=r_i $, therefore from (\ref{F1}) the singular part may be identified as,
\begin{eqnarray}
F_{singular}&=&\frac{1}{2} \left[ \left( \frac{\partial^{2} F}{\partial T^{2}}\right)_{Q}\right]_{r_+=r_i} (T-T_i)^{2}+\frac{1}{2} \left[ \left( \frac{\partial^{2} F}{\partial Q^{2}}\right)_{T}\right]_{r_+=r_i} (Q-Q_i)^{2}\nonumber\\
&=& -\frac{C_Q}{2T_i}(T-T_i)^{2}+\frac{K^{-1}_{T}}{2Q_i} (Q-Q_i)^{2}.
\end{eqnarray} 
Using (\ref{delta}), (\ref{CQ1}), (\ref{KT}) and (\ref{e4}) we finally obtain 
\begin{equation}
F_{singular}= a_i \epsilon^{3/2}+b_i \Pi^{3/2}\label{F}
\end{equation}
where,
\begin{equation}
 a_i=-\frac{A_i T_i }{2}~~~ and, ~~~ b_i=\frac{Q^{1/2}_{i}r_i M^{1/2}_i B_i}{2^{3/2}D^{1/2}_i}.
\end{equation}
Finally, from (\ref{F}) we note that in order to satisfy (\ref{ghf}) we must have $ p=q=2/3 $. Note that although the scaling parameters $ p $ and $ q $ are in general different for a generalized homogeneous function, this is a special case  where both the numbers $ p $ and $ q $ have identical values. In other words $ F $ behaves as a usual homogeneous function. 

In standard thermodynamics, the various critical exponents are related to the scaling parameters as, \cite{stanley1}
\begin{eqnarray}
\alpha = 2-\frac{1}{p},~~~\beta =\frac{1-q}{p},~~~\delta =\frac{q}{1-q}~~\nonumber\\
~~ \gamma =\frac{2q-1}{p},~~~\psi =\frac{1-p}{q},~~~\varphi =\frac{2p-1}{q}
\end{eqnarray}
It is reassuring to note that these relations are also satisfied for BI AdS black holes. One can further see that elimination of the two scaling parameters ($ p $ and $ q $) from the above set of relations eventually leads to (\ref{scaling laws}). 

Finally, we would like to find out the rest of the two critical exponents $ \nu $ and $ \eta $ which are associated with the  correlation length and correlation function respectively. Assuming the additional scaling relations \cite{stanley1},
\begin{equation}
\gamma = \nu (2-\eta),~~~ 2-\alpha = \nu d \label{add}
\end{equation}
to be valid, where $ d(=3) $ is the spatial dimensionality of the system, we find  
\begin{equation}
\nu =1/2,~~~ \eta =1.
\end{equation}
Although for most of the conventional thermodynamic systems with $ d\leq4 $ these relations are found to be satisfied, still it is not clear at the moment whether they are indeed valid in case of gravity theories \cite{cr11}. In this sense the values for $ \nu $ and $ \eta $ given above are more suggestive than definitive. 

\section{Conclusions}
In this paper, based on a standard thermodynamic approach, we have provided a general scheme which could be employed to study the critical phenomena in AdS black holes. While the explicit phase transition points could only be determined from a numerical analysis, the subsequent study and determination of the critical exponents was totally analytical. Using a canonical ensemble, we have exploited this scheme to study the critical behavior in Born-Infeld AdS black holes. Based on this novel approach we have calculated all the static critical exponents ($\alpha=1/2,\beta=1/2,\gamma=1/2,\delta=2,\varphi=1/2,\psi=1/2$) which satisfy the so called  thermodynamic scaling relations near the critical point. Also, we have explored the scaling hypothesis which has been found to be compatible with the scaling relations near the critical point. The scaling parameters have been found to possess identical values ($ p=q=2/3 $), which in general is not the case for a generalized homogeneous function.  Furthermore we have checked the additional scaling relations in order to gain some insights regarding the critical exponents ($ \nu=1/2,\eta=1 $) associated with the spatial correlation. From our analysis it is also evident that in the appropriate limit ($ b\rightarrow\infty $,$ Q \neq 0$) one can recover the critical behavior of thermodynamic functions for RN AdS black holes near the critical point. As a matter of fact we find that the respective values for the critical exponents do not change during this transition from BI AdS to RN AdS black hole. This further suggests the fact that different thermodynamic systems may belong to the same universality class. Thus from our analysis we have been able to justify the similarities between the thermodynamic behavior of BI AdS and RN AdS black holes which were observed earlier \cite{fernando}. Incidentally, the critical exponents $ \alpha $ and $ \gamma $ were obtained earlier \cite{cr8} for the RN AdS case and these agree with our results.

Although we have resolved a number of vexing issues regarding the critical behavior of charged AdS black holes, still there remain a few more issues that admit a further investigation into the subject. For example in order to calculate the critical exponents $ \nu $ and $ \eta $ we have assumed that the additional scaling relations to be valid, which may not be true in practice. Therefore as an alternative approach one should compute them from a knowledge of correlation scalar modes in the BI AdS back ground which would further clarify our calculations. It would also be interesting to exploit the AdS/CFT duality in order to gain a new insight to the subject. Another interesting point in this context is to explore the underlying renormalization group scheme to study the critical phenomena in black holes which could explain the scaling relations in a better way. We want to put all these issues as a future perspective in order to make a further probe into the subject of critical phenomena in black holes. We believe that our approach could illuminate these and related points regarding the underlying microscopic structure of black holes.

{\bf{ Acknowledgement:}}\\
 D.R would like to thank the Council of Scientific and Industrial Research (C. S. I. R), Government of India, for financial help.

  

\end{document}